\documentstyle[preprint,aps]{revtex}
\draft
\begin{document}
\title{
Onset of Vortices in Thin Superconducting Strips and Wires
}
\author{I.~Aranson, M.~Gitterman
and B.~Ya.~Shapiro}
\address{
Department of Physics and Jack and Pearl Resnick Institute of Advanced
Technology, \\ Bar Ilan University, Ramat Gan  52900, Israel
}

\maketitle
\begin{abstract}
Spontaneous nucleation and the consequent penetration of vortices into thin
superconducting films and wires, subjected to a magnetic field,
can be
considered as a nonlinear stage of primary instability of
the current-carrying superconducting state.
The development of the instability
leads to the formation of a chain of vortices in strips
and helicoidal vortex lines in wires.
The boundary of instability was obtained analytically.
The nonlinear stage was investigated by simulations of the
time-dependent generalized Ginzburg-Landau equation.
\end{abstract}


Finite resistance in massive type-II superconductors
subjected to a supercritical magnetic field
$H_{c1}$ is caused by the motion of vortices \cite{Abrik,Gen,gorkov,Ger}.
The order parameter is suppressed at a distance
whose magnitude comparable to the correlation
length  $\xi$  within the core  of the vortex, and
the induced currents and
fields decay at  the penetration length $\lambda$.
The dynamics of vortices both in massive three-dimensional systems
and in two-dimensional films has became the subject of
intensive studies (both analytical \cite{Dors} and numerical \cite{Jap}).

Another limiting case is quasi-one-dimensional systems
(transverse dimensions smaller than $\xi$ and $
\lambda$) such as narrow microstrips and channels.
Their resistance is
caused by phase-slippage centers (PSC), which nucleate  spontaneously
in some narrow region of currents even without
a magnetic field \cite{Langer,Ivlev,Kramer}.

We consider here systems with transverse geometry intermediate
between the two cases described above; i.e., for which
the radius $R$ of 3D wires satisfies the condition:
$\xi \ll R \ll \lambda$. Likewise for strips one assumes that
$h d \ll \lambda^2$ together with $h \gg  \xi$, where $h$
is the thickness and
$d$ is the width of the strip.
We assume that the homogeneous external magnetic field
is applied normal to the strip and
parallel to the wire's axis. The current is applied
lengthwise to the strip and parallel to the wire's axis.
Our aim is to elucidate the onset and penetration
of vortices
into a sample.

The problems of the surface instabilities
and consequent nucleation of the vortices in bulk superconductors
was widely discussed (see, e.g. \cite{Parks}).
The critical magnetic field for bulk semiinfinite
superconducting sample, known as the "superheating field",
was obtained by several authors
(see, e.g. \cite{superh}). However, to our knowledge,
for thin superconducting films and wires with
intermediate geometry ($h d \ll \lambda^2$ or
$\xi \ll R \ll \lambda$) such problems were not studied.

One can expect  that the
stationary, spatially inhomogeneous current-carrying superconducting
state becomes unstable when the applied current $j$ and/or magnetic field $B$
exceed some critical value(s).  Because of the translation invariance along
the longitudial direction,  the perturbations arising
for the order parameter have the form of a standing wave with
finite modulation wavenumber $q$.  In virtue of the inhomogeneity
of the stationary state,  the perturbative eigenfunctions should be
mostly localized near the edges. Exponential growth in time of the
perturbations finally results in the formation of single zeroes
(lines of zeroes for wires)
of the order parameters
on a set
of points placed at the edges of a strip
at a distance $2 \pi /q$ apart.
Such zeroes serve as the  "seeds" for vortices which
will be driven away later
on from the edges by the current and fields. In this article we
present the analytic treatment of this problem. Our results are
supported by the numerical simulations of the TDGLE in the region where
the analytic approach fails.

Although the vortices in thin strips and wires are not
topologically different
from the usual vortices in massive superconductors, the  intermediate
geometry results in significant differences. In contrast to
massive superconductors the magnetic field remains practically homogeneous
on the scale of the strip or wire. As a result, only the vortex
phase rather than the
magnetic flux carried by the
vortex is quantized \cite{foot}.
Moreover, the values of
critical currents and fields obtained
happens to be strongly dependent on the transverse dimension.
The vortices in thin strips and wires are different also from
the $PSC$ nucleating in a very thin wire ($R \sim \xi$ ) where one needs the
potential difference across a sample, and instability typically appears
only in a narrow band of currents \cite{Kramer}. Note that still another
type of vortices occurs in superfluid jets above the critical velocity
\cite{Sonien}, which formally corresponds to zero magnetic field with the
superconducting current greater than the depairing current.

We start from the celebrated time-dependent Ginzburg - Landau
Equation (TDGLE) in the form  \cite{gorkov}

\begin{eqnarray}
u (\partial_t+i\mu)\Psi &=&
\left(\nabla-2i{\bf A}\right)^2\Psi +
\left(1-\left|\Psi\right|^2\right)\Psi ,
\label{eq1_1} \\
{\bf j}&=& \left|\Psi\right|^2\left(\nabla\varphi-2{\bf A}\right)
- \left(\nabla\mu+2\partial_t{\bf A}\right),
\label{eq1_2} \\
\nabla\cdot{\bf j} &=& 0 \;\;\;\;,\;
\nabla\cdot{\bf A} = 0\;,
\label{eq1_5}
\end{eqnarray}
where $\Psi$ is the (complex) order parameter, $\varphi = arg\Psi$,
${\bf A}$ and $\mu$ are vector and scalar potentials, and
${\bf j}$ is the current density.
The value of the
parameter $u$ has to be obtained from microscopic theory \cite{gorkov}.
The usual dimensionless units are used \cite{Kramer}.

Eqs. (\ref{eq1_1}-\ref{eq1_5}) apply
for thin strips and wires. Indeed, the condition $d h \ll \lambda^2$
for strips and $R \ll \lambda$ for wires enable us to neglect the
magnetic field created by currents \cite{pearl,AKW} and omit thereby the
appropriate Maxwell equation.

Let us start with strips.
We choose the origin of the coordinate frame at the mid-point
of the strip with the $x$ axis lengthwise and the $y$ axis
in the lateral direction, so that the edges are located at $(x,-d/2)$ and
$(x,d/
2)$.
The normal to the strip magnetic field $B$ is described by the vector
potential
${\bf A} = (  -B y ,0, 0)$ (see Fig.1 ).

The TDGLE  has a stationary current-carrying
superconducting solution of the
form
\begin{eqnarray}
\Psi(x,y) = F(y) \exp ( i k x) , \;\;\; \mu =0
\label{stat} \\
j(y) = F(y)^2 (k + 2 B y)
\label{current}
\end{eqnarray}
where  $F(y)$ satisfies the following nonlinear equation
\begin{equation}
\partial_y^2 F + \left(1- F^2 -(k + 2 B y)^2\right) F = 0
\label{stat1}
\end{equation}
subject to the no-flux  (superconductor-vacuum)
boundary conditions $\partial_y F(d/2) = \partial_y F(-d/2) = 0$.
This solution  is the only stable superconducting state for $B<B_c$.

For the case of weak magnetic fields
($B \ll k/d$) one can find a perturbative
solution  of Eq. (\ref{stat1}). In lowest order the solutions for
$F(y)$ and the mean current
$j_0 =  \frac{1}{d} \int_{-d/2}^{d/2} j(y) d y$ have the form
\begin{eqnarray}
F = F_0 - \frac{2 k B y} {F_0} +
\frac{ \sqrt{2}  k B \sinh (\sqrt{2} F_0 y)}{F_0^2 \cosh (\sqrt{2} F_0 d/2)}
+O(B^2)
\label{asym3}
\end{eqnarray}
where $j_0 = k (1-k^2)  - k d^2  (B^2 + O(1/d^2))$  and $ F_0 = \sqrt{1-k^2}$
(note that $F(y)-F_0$ is antisymmetric only for $B d \ll 1$).
For arbitrary values of the magnetic field $B$
Eq. (\ref{stat1})
can be solved only numerically. For the numerical solution
of the
nonlinear boundary problem we used the deferred-difference method \cite{NAG}.
The characteristic form of $F(y)$ is shown in Fig. 2.

We seek the perturbative solution of the TDGLE in the form
$\Psi = \left(F(y) + \hat{\zeta}(x,y,t)\right) \exp( i k x)$.
Linearizing TDGLE with respect to
 $\hat{\zeta},\mu$,
splitting real and imaginary parts of $\hat{\zeta} = \hat{a} +
i \hat{b}$,
and representing the solution in the form
$ (\hat{a}, \hat{b}, \mu ) \sim (a(y),b(y),\mu(y)) \exp ( i q x + \omega t)$,
we obtain the following equations for the functions $a(y),  b(y), \mu(y)$
\begin{eqnarray}
\omega  u a & = & \partial_y^2 a + (Z(y,q) -2 F^2) a
-2 i   q (k +2 B y  ) b \nonumber \\
 \omega u  b & = & \partial_y^2 b + Z(y,q) b
+ 2  i   q (k+2 By ) a  - u F \mu         \nonumber \\
F  \omega u i b &=&\partial_y^2  \mu - (q^2+ u F^2)  \mu
\label{per5_0}
\end{eqnarray}
where $Z(y,q)=1 - (k+2 B y )^2 - F^2 -q^2$.
For convenience (in order to deal only with real coefficients in
the Eqs. (\ref{per5_0}) we redefine $b \to i b$. Then the Eqs. (\ref{per5_0})
are of the form
\begin{eqnarray}
\omega  u a & = & \partial_y^2 a + (Z(y,q) -2 F^2) a
-2   q (k +2 B y  ) b \nonumber \\
 \omega u  b & = & \partial_y^2 b + Z(y,q) b
-2   q (k+2 By ) a  - u F \mu \nonumber \\
F  \omega u b &=&\partial_y^2  \mu - (q^2+ u F^2)  \mu
\label{per5}
\end{eqnarray}

Eqs. (\ref{per5}) represent the eigenvalue problem for $\omega$ which has to
be solved for the no-flux boundary conditions for $a,b$ and $\mu$.
The stationary solution becomes unstable  when for the first time
the negative $\omega(q)$ achieves zero.
For $\omega=0$
the equation for $\mu$ has no solutions
obeying the boundary condition. Therefore,  we are left with two equations
for $a$ and $b$  ($\mu=0$)
\begin{eqnarray}
 \partial_y^2 a + (Z(y,q) -2 F^2) a
-2   q (k +2 B y  ) b &=& 0  \nonumber \\
 \partial_y^2 b + Z(y,q) b
-2   q (k+2 By ) a &=&0
\label{per5_1}
\end{eqnarray}

The system (\ref{per5_1}) in general requires numerical solution. These
equations
contain four control parameters,
namely, the  magnetic field $B$,
the mean current $j_0$ which is determined
by $k$,
the width
of the strip $d$ and the modulation wavenumber $q$.
The numerical solution was
implemented in the following way:
1) The stationary solution $F(y)$ was determined (numerically) and
this was then substituted into Eqs. (\ref{per5_1});
2) For  fixed $q$,
Eqs. (\ref{per5_1}) were solved
for two sets of boundary conditions at the lower edge:
$\partial_y a_1 (-d/2) = \partial_y b_1 (-d/2) =0,
 a_1 (-d/2) = 1, b_1 (-d/2) = 0$ and
$\partial_y a_2 (-d/2) = \partial_y b_2 (-d/2) =0,
a_2 (-d/2) = 0, b_2 (-d/2) = 1$; 3) Then
the solutions at the second edge were examined;
4) The eigenvalue $q$ was obtained from the
condition
\begin{equation}
\partial_y a_1 \partial_y b_2 -\partial_y a_2 \partial_y b_1 = 0.
\label{boncon}
\end{equation}

The critical curves for the onset of instability in the
$(j,B)$ plane for several widths of the strip
$d$ are depicted on Fig. 3. The eigenfunctions
$ a(y), b(y)$ are shown in Fig. 2.
It can be seen that for $B \ne 0$ the eigenfunctions
are strongly localized near the edge of the strip where the suppression of
the order parameter is maximal.  The critical curves are  terminated at the
depairing current $j_p = 2/ \sqrt27 \approx 0.387 $, which
corresponds to $k_p=1/\sqrt3$. In this region a
very weak magnetic field is needed to destroy superconductivity.
In the practically relevant
limit $d \gg 1$ (well-separated edges), Eqs. (\ref{per5_1})  can be essentially
simplified by
adiabatic elimination of $a(y)$ from the second Eqs.
(\ref{per5_1}) resulting in:
\begin{equation}
\partial_y^2 b = q^2 (1 - \frac{4 (k + 2 By)^2}{2 - 2 (k + 2 By)^2 +q^2}) b
 + O(1/d^2).
\label{pern}
\end{equation}
For $k \to k_p$ also $q \to 0$ and the critical curve is given
by the solvability condition
$\int_{-d/2}^{d/2} (1 - \frac{4 (k + 2 By)^2}{2 - 2 (k + 2 By)^2 }) d y =0 $.
The straightforward calculations results in  the following conditions:
$j_p-j = k_p B^2 d^2$ and $B^2 d^2 = - 3 \delta/ k_p$
where $\delta = k - k_p \ll 1$.
However this trivial solution is relevant only for very small
$\delta  \sim 1/d^2$. For larger $\delta $ this curve merges with
another curve which turns out to be unstable for smaller $B$.
This curve obeys the scaling $(q^2 d,  \delta d, B d^2) \sim O(1)$.
For this case the Eq.(\ref{pern}) is reduced to the Airy equation:
\begin{equation}
\partial_y^2 b = (\alpha+ \beta y)  b
\label{Airy}
\end{equation}
where $\alpha=\frac{3}{16} q^2 (-16 \sqrt3 \delta (1-q^2)
 + 4  q^2 -72 \delta^2 -3 q^4), \beta = -6 \sqrt3 q^2 B$
(these expressions  were obtained with the aid of  the Maple program
for analytical calculations).
The critical curve for the Eq. (\ref{Airy}) is given by the
conditions (cf. (\ref{boncon}))
\begin{eqnarray}
\partial_y Ai(\gamma_1) \partial_y Bi(\gamma_2)
-\partial_y Bi(\gamma_1) \partial_y Ai(\gamma_2)=0 \nonumber \\
\frac{\partial}{\partial q } \left[
\partial_y Ai(\gamma_1) \partial_y Bi(\gamma_2)
-\partial_y Bi(\gamma_1) \partial_y Ai(\gamma_2) \right]=0
\label{boncon1}
\end{eqnarray}
where $\gamma_{1,2}=\beta^{1/3} (\mp d/2 + \alpha/\beta)$.
The critical curves obtained by solution of
Eq. (\ref{pern}) are shown also on Fig. 3.
The critical curves obtained from the Airy equation (\ref{Airy},\ref{boncon1})
coincide
with those within the thickness of the line.
We conclude  that the "$d \gg 1$" approximation works extremely well.
As can be seen from Fig. 3, the critical field $B_c$ decreases with $d$, and,
finally  $B_c$ vanishes  for $d \to \infty$.
This is consistent with
well-known results \cite{Abrik} that the critical magnetic field in an infinite
film is
always zero.
Figures 2-3 support the physical picture of an appearance of
a chain of vortices at one edge of the strip (where the eigenfunctions
are localized). It is useful to mention that
the critical curves appear to be insensitive to
the particular properties of the superconductor given by the material
constant $u$.

In order to follow the consequent stages  of the instability, which cannot be
covered by the linear analysis,
we performed direct numerical
simulations of TDGLE (with $u=5.79$ \cite{gorkov}).
The initial conditions
were chosen typically as a homogeneous superconducting state perturbed
by  small amplitude noise.
We used no-flux boundary
conditions in the transverse direction ($\partial_y \Psi =0$) and normal metal-
superconductor boundary conditions in the longitudial direction
($\Psi(x,y) \to 0$ for $x \to 0, L$, where $L$ is the strip length).
The numerical scheme was the generalization of
the split-step method described in \cite{AGS,AKW,weber}, the number of the
grid points was 256$\times$128  and the timestep was $0.05-0.1$.
Results of the simulations, shown in Fig. 3, 4
clearly agree with the stability analysis.

The analytical solution and the numerical simulations of TDGLE clarify
all stages of the process: initial development of small perturbations
due to applied current and/or  field (Fig. 4a),
their  localization at some points at the lower edge (Fig. 4b) and an
appearance of the chain of single vortices and its consequent tearing off and
propagation inside
the strip (Fig. 4c).
Then the process  repeats itself.
When the vortex is sufficiently far away from
the edges, its motion can be described, in principle,
by a simple first order equation
\cite{AGS,AKW}. This periodic nucleation of vortices at  the edge
was observed for sufficiently
wide range of transport currents $0<j<j_p$. Thus, one
has a dynamic resistive state near the critical line. A gradual
increas of
the field $B$ leads to faster nucleation, also the vortex motion itself
looses the regularity and reminds of some sort of vortex turbulence
\cite{AGS,Jap}.
In that case one has also nonperiodic oscillations of the
voltage across the sample, as it was observed in numerical simulations
\cite{AGS,Jap}. Very close to  the critical current $j_p$ the
film doesn't exhibit dynamic resistive states  and  undergoes direct transition
to normal state. Alternatively, for $j \to 0$ the period of the nucleation
diverges.

Our analysis can be generalized for the case of superconducting wires
when  ${\bf B} $ is parallel to the axis $z$ of wire. Then the
potential can be chosen in the form
${\bf A}= (A_r, A_\theta , A_z)= (0, B r/ 2 , 0)$.

The equation for the stationary order parameter has the form
(compare with Eq. (\ref{stat1}))
\begin{equation}
\partial_r^2 F  + \frac{1}{r}\partial_r F+  \left(1- F^2 -k^2 - ( B r)^2\right)
F = 0
\label{stat2}
\end{equation}
Eq. (\ref{stat2}) is subjected to the boundary conditions
$\partial_r F(0) = \partial_r F(R) =0$.

We seek a perturbative solution in the form
$\Psi = \left(F(r) + \zeta(r, \theta,z,t) \right) \exp (i k z)$
Notice, that unlike Eqs. (\ref{per5}), one now has the full
3D problem. In contrast to strips the wires also possess
rotation symmetry. Therefore, the perturbative solution should
break both translation and rotation symmetry, creating, thereby,
helicoidal structure.

Linearizing now the TDGLE, and substituting the
perturbations in the form $(a, b, \mu ) \sim \exp (i n \theta +  i q z + \omega
t)
$
($a,b$ have the same meaning as for a strip),
we are left for the critical value
$\omega=0$ with  the following equations
\begin{eqnarray}
\Delta_n a_n +(Z_1(r,q) -2 F^2)
 a_n
-2   q k b_n +  2 B n b_n   = 0 \nonumber  \\
\Delta_n b_n
+ Z_1(r,q) b_n
-2   q k a_n + 2 B n  a_n  = 0
\label{per11}
\end{eqnarray}
where $Z_1(r,q)= 1 - k^2 - (B r)^2 - F^2 -q^2,
\Delta_n = \partial^2_r + \frac{1}{r}  \partial_r  - \frac{n^2}{r^2}$.

The boundary conditions
 read  $\partial_r a_n = \partial_r b_n = 0$ for $r = R$
and $a_n,b_n \sim r^n$ for $r \to 0, n \ne 0$. For $n=0$ the conditions
translate to $\partial_r a_0(0) = \partial_r b_0 (0)  = 0 $.
The numerical solution of  Eqs. (\ref{per11}) is similar to that of
the strip if the fact that they
contain the rotation number $n$ is
taken into account.
Critical curves  for each $n$ intersect in some
complicated way on the $(j,B)$ plane.
In order to select the
most unstable rotation mode we used the following algorithm. We
reproduced the critical curves for the 6 first modes ($n=0 .. 5 $,
other modes turned
out to be unimportant for the
chosen radius of the wire) as a function of $k$, and
then
for each given $k$ we selected the mode which is  unstable for lowest value
of $B$. As a result, the critical curve in the $(j,B)$ plane
consists of several pieces of different
$n$. As can be seen from Fig. 5 the rotation symmetry changes with
$B$, namely, for small $B$
the axi-symmetric mode $n=0$ appears while for larger
$B$ the higher helocoid solutions occur.
The number of the "important" modes grows with the radius of the wire.
For $R \gg 1$ the critical curves can be described by the reduction
to effective Airy equation as that for the strip.
For macroscopic wires the helicoid structures are well-known (see, e.g.,
\cite{Brandt}).

It is a challenging problem to generalize the above
formalism to a magnetic field orthogonal to the
axis of wire. However, the stationary state is then described by a
2D nonlinear
partial differential equation, and
the analysis is very hard. One can speculate that the critical curves for this
case will not
much differ from that of a strip with $d \approx 2 R$.

Finally, we have described all the stages of the
appearance and the propagation of a chain of vortices in thin strips
and wires, in contrast to the usual approach where
the existence of vortices is assumed {\it ad hoc}.
All existing methods of experimental observation
of fluxon dynamics, e.g., voltage-current characteristics, temporal
pulses at a measured voltage, etc.  may be used  to check our
analysis. High-temperature superconductors are most suitable for these
experiments. Indeed, they have large values of $\lambda$ and, hence,
samples can be prepared satisfying the conditions $d h \ll \lambda^2$
and $R \ll \lambda$.
In particular, for the $YBaCuO$ system with parameters
$\lambda = 1500$ \AA,~$d = 500$ \AA ~and $h = 10$ \AA,
one has $ d h / \lambda^{2}   \ll 1$.
For this system the scale of physical units shown in Fig. 2-5 are:
$y_{phys} \sim y \times 10 \AA$, $B_{phys} \sim B \times 5 \cdot 10^3$ Gs and
$j_{phys} = j\times 2.5 \cdot 10^8$ $A/cm^2$.  The characteristic
time scale is given by $t_{phys}= t \times 10^{-9}$ sec.

The authors are grateful to L. Kramer, A.I. Larkin and N.B. Kopnin for
fruitful discussions.
The work of I.A. and B.Y. S. was supported in part by the Rashi Foundation.
The support of the Israeli Ministry of Science and
Technology is kindly acknowledged.

\references{
\bibitem[\dag]{foot} Such a vortex is often called a fluxoid (see \cite{Gen}).
\bibitem{Abrik} A.A. Abrikosov, {\it Fundamentals of the Theory of Metals},
(Elsevier, New York,  1988).

\bibitem{Gen} P. G. de Gennes, {\it Superconductivity of Metals
and Alloys}, (Addison-Wesley, Redwood City, 1989).

\bibitem{gorkov}  L.P. Gor'kov
and N.B. Kopnin, Usp. Fiz. Nauk, {\bf 116}, 413 (1975).

\bibitem{Ger} G. Blatter, M.V. Feigelman, V.B. Geshkenbein,
A.I. Larkin, and V.M. Vinokur, Vortices in High Temperature Superconductors,
Rev. Mod. Phys, 1994 (to be published).

\bibitem{Dors} F. Liu, M. Mondello, and N. Goldenfeld, Phys. Rev. Lett.,
{\bf 66}, 3071 (1991); H. Frahm, S. Ullah,  and A. Dorsey, Phys. Rev. Lett.,
{\bf 66}, 3067 (1991).
\bibitem{Jap}
A. Shinozaki and Y. Oono, Phys. Rev. Lett, {\bf 66}, 173 (1991);
R. Kato, Y. Enomoto, and S. Maekawa, Phys. Rev. B, {\bf 44}, 6916
(1991); {\bf 47}, 8016 (1993); M. Machida and Kaburaki,
Phys. Rev. Lett. {\bf 71}, 3206 (1993).

\bibitem{Langer} J.S. Langer, Rev. Mod. Phys., {\bf 52}, 1 (1980).
\bibitem{Ivlev} B.I. Ivlev and N.B. Kopnin, Adv. Phys, {\bf 33}, 47 (1984).
\bibitem{Kramer} R.J. Watts-Tobin, Y. Kr\"ahenb\"uhl,  and L. Kramer,
J. of Low. Temp. Phys., {\bf 42}, 459 (1981); L. Kramer and R. J. Watts-Tobin,
Phys. Rev. Lett, {\bf 40}, 1041 (1978).
\bibitem{Parks} Y.V. Sharvin, Zh. E.T.F. -Pi'sma, { \bf 2}, 287 (1965)
[translation: JETP Letters {\bf 2}, 183 (1965)];
B. S. Chandrasekhar, D. E. Farrell, and S. Huang, \prl, {\bf 18}, 43 (1967);
B. L. Brandt and R. D. Parks, \prl, {\bf 19}, 163 (1967).

\bibitem{superh}
V.L. Ginzburg, Zh. E.T.F., {\bf 34}, 113 (1958)
[translation: JETP {\bf 7}, 78 (1958)];
 L. Kramer, Phys. Rev. {\bf 170}, 475 (1968);
Z. Physik, {\bf 259}, 333 (1973).

\bibitem{Sonien} P.I. Soininen and N.B. Kopnin, \prb, {\bf 49}, 12087 (1994).
\bibitem{pearl} J. Pearl, Appl. Phys. Lett, {\bf 5}, 65 (1966).
\bibitem{NAG} NAG Fortran Library, Mark 15.
\bibitem{AGS} I. Aranson, M. Gitterman, and B. Y. Shapiro, Motion of
vortices in thin superconducting films, to appear in
J. of Low Temp. Phys., 1994.
\bibitem{AKW} I. Aranson, L. Kramer, and A. Weber, J. Low. Temp. Phys, {\bf
89},
859 (1992).
\bibitem{weber} A. Weber and L. Kramer, J. of Low  Temp. Phys., {\bf 84},
289 (1991).
\bibitem{Brandt} E.H. Brandt, Phys. Rev. Lett., {\bf 69}, 1105 (1992).
}

\begin{figure}
\caption{Coordinate frame for the superconducting strip.
\label{fig1}}
\end{figure}

\begin{figure}
\caption{The amplitude of order parameter $F(y)$ and the perturbative
solutions $a(y),b(y)$ as functions of $y$
obtained from numerical solutions of Eq. (\protect{\ref{stat1}}) and
Eqs. (\protect{\ref{per5}})
for $q=0.2804$,
$d=30, k=0.3, B=0.014$ and $j=0.22$.  }
\end{figure}

\begin{figure}
\caption{The critical $q_c$ and $B_c$ versus $j$
for two different values of $d$.
The diamonds and heavy dots show the loci of stable and
unstable stationary solutions
of TDGLE obtained  by numerical simulations.
Dashed  lines show the critical curves are given by Eq. (\protect \ref{pern}).
}
\end{figure}

\begin{figure}
\caption{
Results of numerical simulations of the TDGLE. The time runs
from the top to the bottom:
a) $t=90$, b) $t=150$ and c) $t=210$. The  parameters used are
$d=30, B=0.018, j=0.2$
and the length of the strip $L=70$. The current is applied along
$x$-axis and the magnetic field is perpendicular to the strip.
On the gray coded  snapshots of $|\Psi(x,y)|$ (
$|\Psi(x,y)|=0$ is shown in black and $|\Psi(x,y)|=1$ is shown in white)
one clearly sees the onset and propagation of vortices.
 }
\end{figure}

\begin{figure}
\caption{The $(j,B)$ critical curve for a wire with $R=10$}
\end{figure}

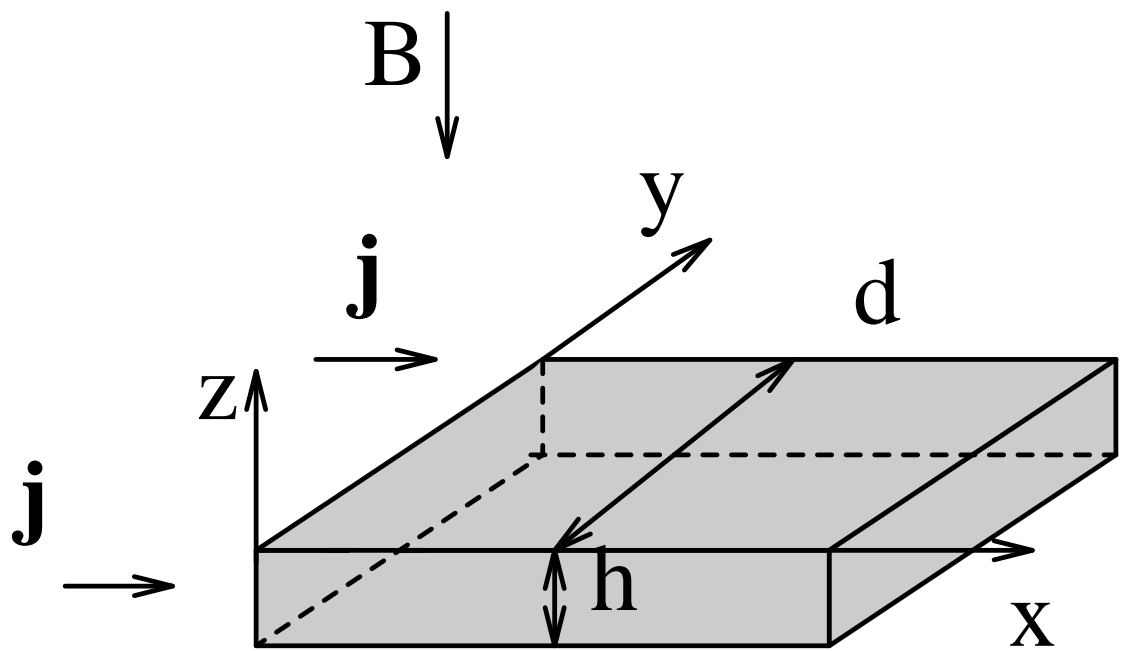
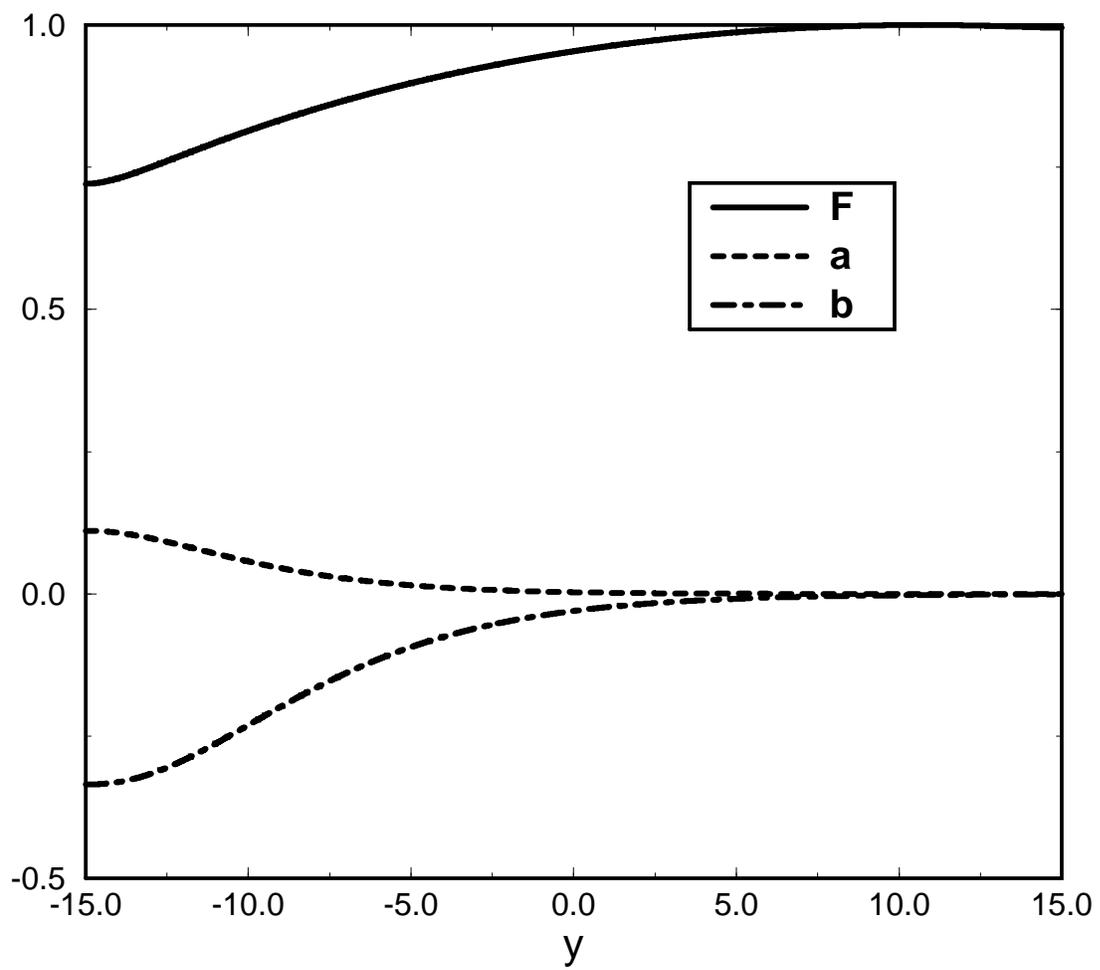
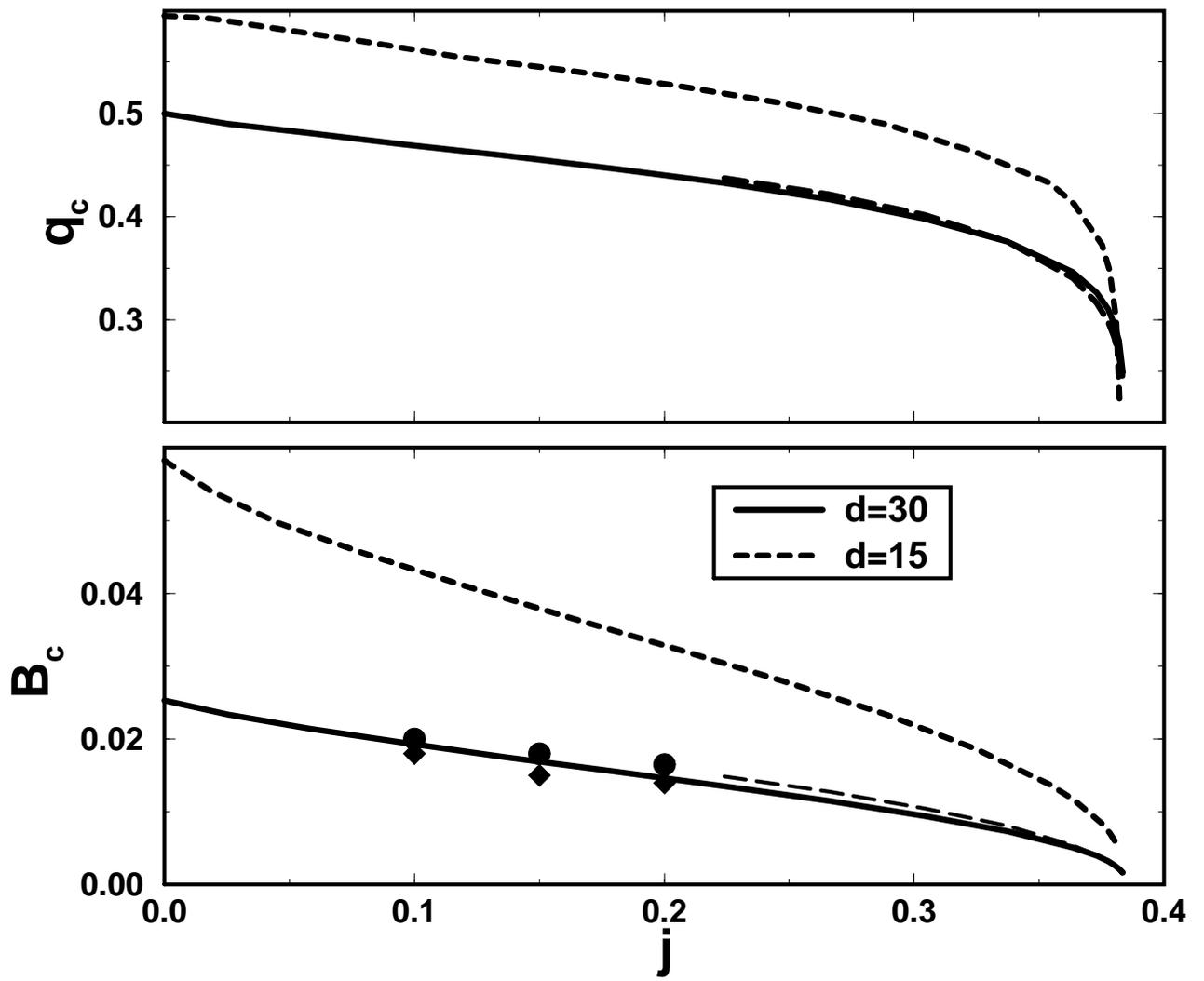
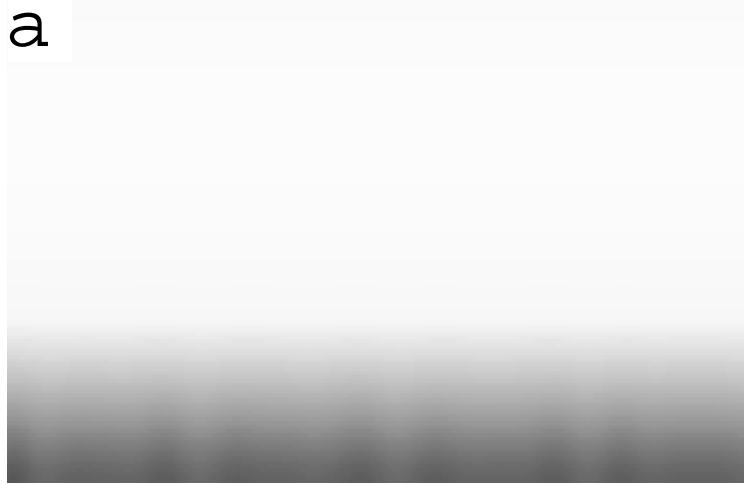
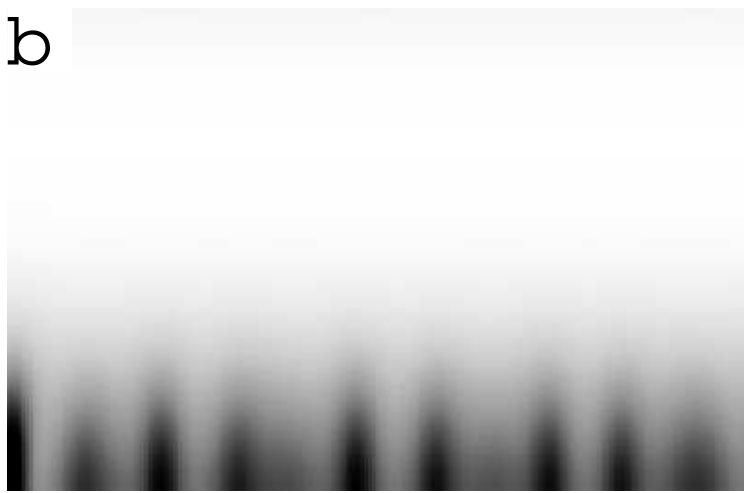
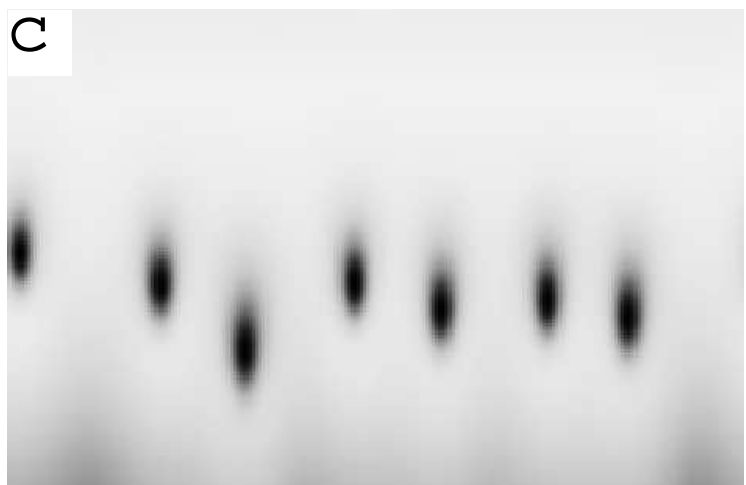
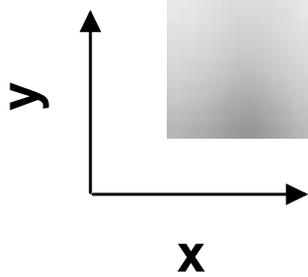
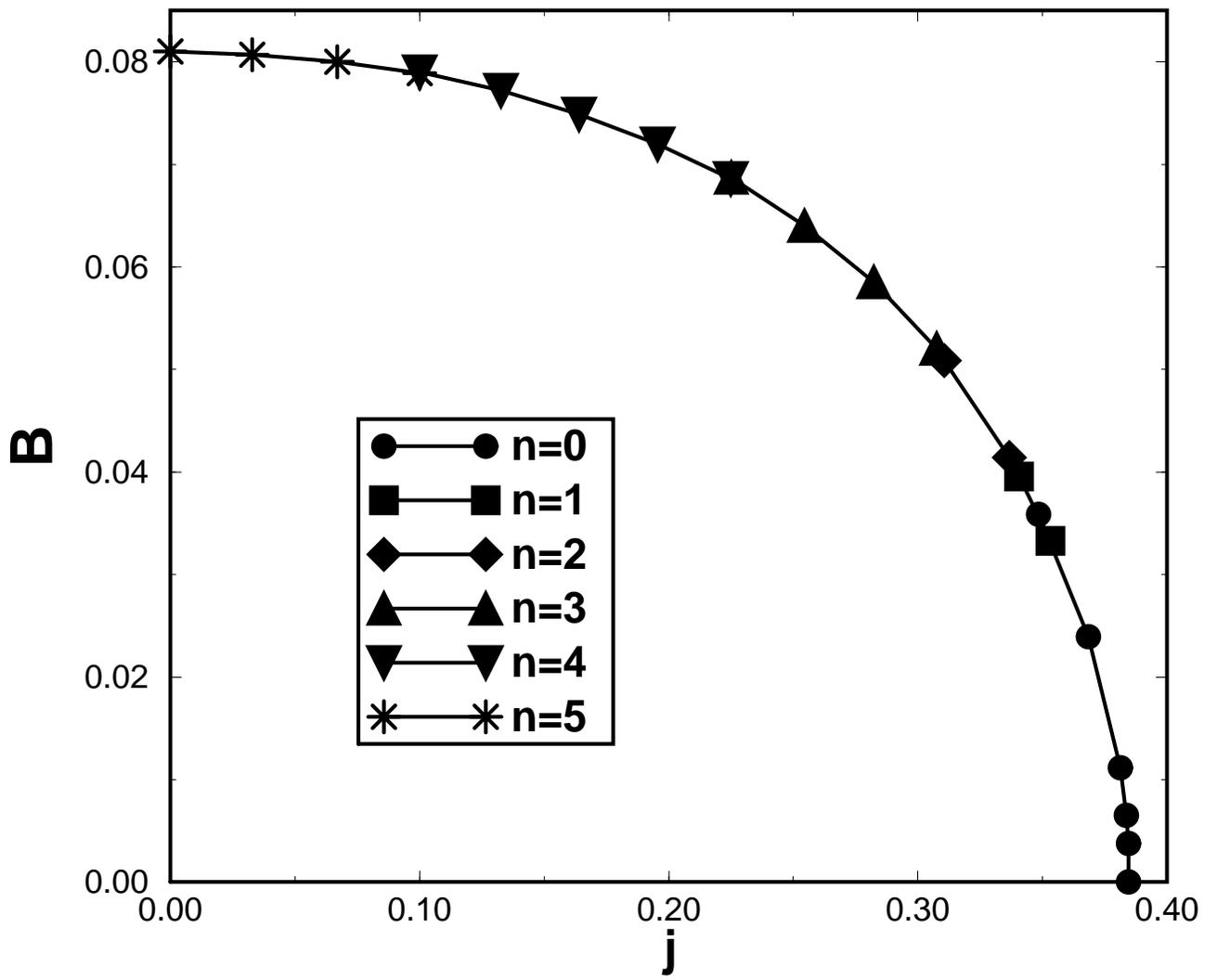
\end{document}